# High-k gadolinium scandate on Si obtained by high pressure sputtering from metal targets and in-situ plasma oxidation

M A Pampillón[1], E San Andrés[1], P C Feijoo[2] and J L G Fierro[3]

[1] Dpto. Física Aplicada III (Electricidad y Electrónica), Universidad Complutense de Madrid, Fac. de CC. Físicas. Av/Complutense S/N, E-28040 Madrid, Spain

2 Dept. d'Enginyeria Electrònica, Escola d'Enginyeria, Universitat Autònoma de Barcelona, Campus UAB, E-08193 Bellaterra, Spain

[3] Instituto de Catálisis y Petroleoquímica, Consejo Superior de Investigaciones Científicas, C/Marie Curie 2, E-28049 Cantoblanco, Spain

**Abstract**

This article studies the physical and electrical behavior of $Gd_{2-x}Sc_xO_3$ layers grown by high pressure sputtering from metallic Gd and Sc targets. The aim is to obtain a high permittivity dielectric for microelectronic applications. The films were obtained by the deposition of a metallic nanolaminate of Gd and Sc alternating layers, which is afterwards in-situ oxidized by plasma. The oxide films obtained were close to stoichiometry, amorphous and with minimal interfacial regrowth. By fabricating metal–insulator–semiconductor capacitors we found that a moderate temperature annealing is needed to enhance permittivity, which reaches a high value of 32 while keeping moderate leakage. Finally, the feasibility of interface scavenging in this material with Ti gate electrodes is also demonstrated.

## 1. Introduction

To continue with the advance in the field of gate dielectrics for metal–oxide–semiconductor field-effect-transistors (MOSFET), it is mandatory to keep up increasing the gate capacitance per unit area. This trend caused the substitution of SiON by hafnium-based dielectrics in 2008 [1, 2]. The main limitation of hafnium oxide is its moderate crystallization temperature [3]. Gadolinium scandate is a promising candidate for these high-k applications due to its favorable properties, such as a high relative permittivity (between 20 and 30, depending on the lattice direction and the composition) [4–6], a large bandgap (higher than 5 eV) [7–9], large conduction and valence band offsets to Si (around 2–2.5 eV) [8] and an excellent thermodynamic stability on Si (up to 1000 °C) [10, 11].

In this work we have used high pressure sputtering (HPS) to deposit the gadolinium scandate, a non-conventional technique initially developed by Poppe et al [12] for the epitaxial growth of high-$T_c$ superconductors. On the other hand, the industry has converged on atomic layer deposition (ALD) as the workhorse to grow the high-k dielectric of the gate stack. The main reasons of this predominance are the sub-monolayer control of thickness, film composition control, conformal deposition, uniformity and repeatability [13]. However, there are some limitations that give room for alternative techniques. For instance, the precursor used in ALD (typically chlorides or organometallics) can cause film contamination from Cl, C and H [14]. Since the deposition is performed at moderate temperatures (around 350 °C [15]), it is hard to avoid completely these elements. On each cycle a very high gas flow is needed to achieve a rapid saturation of the growing surface, implying a high amount of precursors. The chemisorption of the precursors is needed, so some surfaces have to be chemically prepared before ALD (for instance, a thin $SiO_x$ layer is usually used [16]). Finally, there are not yet recipes to grow every material with electronic interest (for instance, Li, Au or Cr [15]).

Sputtering is a technique that can circumvent some of these limitations. The targets can be fabricated with extremely high purity, and since the gas used is typically composed of high purity Ar, $N_2$ and $O_2$, film contamination can be minimal. Non-elementary materials can be obtained with composite targets [17], by co-sputtering [18] or reactive sputtering [19]. Also a nanolaminate can be obtained by sequential sputtering. The growth rate is fairly lineal, and since surface chemisorption is not needed, sputtering is less sensitive to substrate composition. The target material use is much better, and at the end of the target's life, the remaining material can be recycled (for expensive materials this is a great financial advantage). Reverse sputtering can be used to clean in-situ the substrate before deposition. Finally, the deposition temperature can be lower than in ALD [20], so the interfacial layer regrowth can be minimized.

However, conventional sputtering has some drawbacks. The main problem is that extended plasmas can damage the surface. At typical pressures ($10^{-2}$–$10^{-3}$ mbar), the mean free path of the plasma species is in the order of some cm [21]. HPS aims at minimizing this problem by increasing substantially the working pressure, to the 0.5–5 mbar range, where the mean free path is in the $10^{-2}$ cm range. At these pressures the energetic sputtered species suffer many collisions in a short distance, decaying to the thermal energy, and from then they move to the substrate by a diffusion process. This way surface bombardment is minimized, which is key to achieve a well behaved high-k/semiconductor interface.

In our group we deposited previously $Gd_{2-x}Sc_xO_3$ by HPS but using oxide targets, either from a $GdScO_3$ target [17] or by annealing $Gd_2O_3$–$Sc_2O_3$ nanolaminates [22]. In both approaches during the first stages of growth the bare Si surface was exposed to the oxygen present in the plasma that comes from the oxide target. Thus some interfacial oxidation was unavoidable. In this work we have used a different deposition procedure: instead of using oxide targets, we have deposited first a metallic film on Si. This layer is in-situ plasma oxidized afterwards. The goal is to fully oxidize the metallic film but minimizing interlayer regrowth by adjusting plasma oxidation conditions. We developed this oxidation process to obtain $Gd_2O_3$ films in [23]. However, in HPS to obtain compounds from pure targets the only possibility is to produce the intermixing of a sub-nm alternating stack of the elements (in other words, co-sputtering is not possible at the high working pressure of the HPS system). For the depositions of this work, two targets of Gd and Sc were used, and both were radiofrequency (rf) excited continuously during the deposition process. The gadolinium scandate composition can be adjusted then by controlling the relative thickness of the Gd and Sc films.

Another concept that is gaining momentum in the high-k field is the interface scavenging process. It consists on the use of optimized metal electrodes (scavengers) and tailored annealing treatments to achieve interface reduction with minimal impact on high-k quality. This process was initially demonstrated by Kim et al [24] and afterwards has been researched thoroughly, predominantly focusing on $HfO_2$ [25, 26]. This process holds the promise of extremely low EOTs by controlling the interfacial layer thickness and oxygen content. In [27, 28] we showed that this scavenging also works on $Gd_2O_3$ films deposited by HPS on Si and even on InP. In this work we will use Ti/Pt gate electrodes to study the possibility of scavenge the interface of HPS deposited $Gd_{2-x}Sc_xO_3$.

This paper is organized as follows: first a physical characterization of the films was performed, including x-ray photoemission spectroscopy (XPS) to obtain the scandate composition, Fourier transform infrared spectroscopy (FTIR) to study the interfacial layer presence, high resolution transmission electron microscopy (HRTEM) to measure the thicknesses of the films, and grazing incidence x-ray diffraction (GIXRD) to check the crystalline structure. Also, MIS devices were fabricated with pure Pt (to study the plain properties of the dielectric) and Ti/Pt gate electrodes (to assess the potential advantages of scavenging). The evolution of the electrical characteristics with forming gas annealings (FGA) temperature was studied.

2. **Experimental methods**

$Gd_{2-x}Sc_xO_3$ films were fabricated by means of HPS from metallic Gd and Sc targets. First, thin Gd and Sc layers were alternatively deposited in a pure Ar atmosphere at 30 W of rf power and a pressure of 0.5 mbar (base pressure was $2 \times 10^{-6}$ mbar). To achieve this sequential deposition without breaking the vacuum of the chamber, the targets are mounted on a mechanized arm, which places each target above the substrate the time needed to obtain the desired thickness, 4.8 s for Gd, and 14.4 s for Sc. All targets are sputtered simultaneously by separate rf power sources (model PFG 300 from Hüttinger). The goal was to obtain a $Gd_{2-x}Sc_xO_3$ close to stoichiometry. The difference in deposition time comes from the measured different growth rates of Gd and Sc, that were empirically measured for these HPS conditions in previous experiments. In those experiments we measured that these sputtering times and power resulted in ~0.35 nm oxide films for each step. This cycle was repeated 7 times in order to obtain a ~5 nm thick $Gd_{2-x}Sc_xO_3$. Also, pure Gd and Sc layers were grown as reference samples. To have the possibility of fine tuning of the ternary composition (Kittl et al showed in [5] a dependence of permittivity with composition), it is very important that each metallic layer of the nanolaminate (Gd or Sc) is not contaminated by the other material. This could be achieved by turning on and off each target sequentially. However, doing this would decrease throughput, since the plasma needs some stabilization time. Thus we kept both targets working during the whole process instead. No relevant differences in the emission of the Gd or Sc plasmas were detected by glow discharge optical spectroscopy (GDOS) when the other target was ignited. This is possible in HPS since the high pressure (0.5 mbar) confines the plasma to a small vicinity of the target (less than 2 cm) so it is not affected by the other target plasma, which is at a distance of 15 cm. Therefore, it is possible to deposit a nanolaminate of metallic Gd and Sc layers in an Ar atmosphere using HPS, and each step is free of contamination from the other target, even when both targets are being sputtered simultaneously.

Afterwards, a plasma oxidation was carried out in-situ during 100 s in a mixed 95% Ar/5% $O_2$ plasma using the Gd target at 20 W. This process was identical for all samples. The total pressure was 0.5 mbar. In [29] we found that the plasma oxidation was possible with both Gd and Sc targets. We chose the Gd target for the plasma oxidation of $Gd_{2-x}Sc_xO_3$ because we studied it widely [23], obtaining good and reliable results. Both processes (the metal deposition and the plasma oxidation) were carried out at room temperature.

Besides, in order to produce the ternary material from the nanolaminates and gain insight on the thermal behavior, a soak annealing was consecutively performed at temperatures from 300 °C to 700 °C in a forming gas atmosphere (10% $H_2$, 90% $N_2$ composition). For temperatures up to 500 °C the annealing time was 20 min, while for higher temperatures the duration was reduced to 5 min to avoid problems with the Al of the back contact. The RTA furnace used was a RTP-600 from Modular Process Technology.

The substrates used for physical characterization of the high-k films were high resistivity n-Si (100) 2′ wafers, with a resistivity of 200–1000 Ω cm and polished on both sides. On the other hand, for electrical characterization of the films, metal–insulator–semiconductor (MIS) devices were fabricated on single side polished (100) n-Si wafers with a lower resistivity in the 1.5–5.0 Ω cm range. On the substrates used for MIS fabrication, square openings were defined on e-beam deposited field oxide. The device active area ranged from 500 μm × 500 μm to 100 μm × 100 μm. All wafers were cleaned by a standard radio corporation of America process [30]. Just before the introduction to the HPS chamber the Si wafers were etched in a 1:50 HF solution during 30 s to remove the native oxide. After high-k deposition the gate electrode was e-beam evaporated and defined by a lift-off procedure. Pure Pt electrodes 25 nm thick and 5 nm Ti/25 nm Pt stacks were studied as the top electrode. Backside contact was also e-beam evaporated (100 nm thick Ti capped by 200 nm of Al).

Several physical characterization techniques were used: the bonding structure of the dielectric films was analyzed by FTIR spectroscopy using a Nicolet Magna-IR 750 Series II in transmission mode at normal incidence. The wavenumber range measured was 400–4000 cm$^{-1}$. The dielectric film absorbance spectrum was obtained after subtraction of a bareSi substrate spectrum; XPS spectra were obtained by a VG Escalab 200 R spectrometer equipped with a Mg K$_\alpha$ x-ray source (hν = 1254.6 eV), powered at 120 W; TEM lamellas were prepared by manual grinding and polishing, while high resolution TEM images were taken with a Jeol JEM 3000 F at an energy of 300 keV. XRD diffractograms were measured with a PANalytical diffractometer X'Pert PRO MRD, using the Cu K$_\alpha$ line of 0.1541 nm.

Concerning electrical characterization, we measured the high frequency C–$V_{gate}$ and G–$V_{gate}$ curves of the MIS devices using an Agilent 4294 A impedance analyzer at a frequency range between 1 kHz and 1 MHz. The leakage current density (J–$V_{gate}$) was measured using a Keithley 2636 A system. The samples were measured before and after FGA. We extracted parameters such as the equivalent oxide thickness (EOT) with the Hauser's algorithm [31], the flatband voltage ($V_{FB}$) or the interface trap density ($D_{it}$) with the conductance method [32]. This way, we analyzed the influence of the FGA on the dielectric permittivity and the electrical performance.

3. **Results and discussion**

The chemical composition of the samples was measured by means of XPS. No sputtering during the XPS measurement was available, so the information comes from the layer surface.

Pure $Gd_2O_3$ and $Sc_2O_3$ samples together with a $Gd_{2-x}Sc_xO_3$ film were analyzed after the FGA at 600 °C. The scandate was grown aiming at the stoichiometric composition, which is the most widely studied and has demonstrated the k enhancement as compared to the binary constituents [5]. It is important to remember that even the pure binary oxide films were deposited with both targets switched on simultaneously. Also in all samples the in-situ plasma oxidation was performed with the Gd target. The XPS wide scan of survey spectra obtained for these films is shown in figure 1. For all films at 284.8 eV, the C 1s peak shows the presence of adventitious carbon during the measurement. Also, it can be seen that the pure $Gd_2O_3$ sample does not present any trace of Sc, in agreement with the GDOS results (not shown) that pointed out to a Sc-free plasma on the target. In this sample the most relevant Gd binding energy is located at ∼140 eV and arises from the $4d_{5/2}$ level. Also the Gd $4p_{3/2}$ is present at 272.4 eV. On the other hand, in the case of pure $Sc_2O_3$, the most intense feature is located at ∼400 eV and is due to the Sc $2p_{3/2}$ level. Also a slight peak corresponding to Gd is observed, possibly due to the oxidation process that was carried out with the Gd target: XPS detects surface atoms, and a minimal amount of Gd can be sputtered during oxidation. In [28] we concluded that the extraction of Gd in a mixed Ar/O$_2$ atmosphere was small. These XPS results indicate that a low amount of Gd is being sputtered during oxidation. Focusing on the ternary material, both Gd 4d$_{5/2}$ and Sc 2p$_{3/2}$ are observed in the grown layer, indicating a ternary composition. Besides, all the spectra present the O 1s peak.

In order to calculate the chemical composition of the Gd$_{2-x}$Sc$_x$O$_3$ layer, high resolution spectrum of the ternary material sample was obtained, as it is presented in figure 2. The spin–orbit splitting of the Gd 4d core-level (with components 4d$_{5/2}$ and 4d$_{3/2}$ located at around 142.2 and 149.1 eV, respectively) are in agreement with reported Gd$_2$O$_3$ [33, 34]. The doublet for the Sc 2p level at about 401.5 eV (for Sc 2p$_{3/2}$) and at 406.9 eV (for Sc 2p$_{1/2}$) corresponds with Sc$_2$O$_3$ as it is stated in other works [35, 36]. In the case of O 1s, it is observed an asymmetric peak that is fitted with two components: one lower at 529.6 eV and other higher at 531.5 eV. The first one can be related to Sc–O or Gd–O bonds [34, 37], while the position of the second points to Gd–O bonds [32, 38] or OH⁻ groups [39]. Since in pure Gd$_2$O$_3$ the only peak that was found was the one at 531.5 eV (not shown), we can assume that the 529.6 eV peak comes from oxygen—scandium bonds, while the one at 531.5 is due to oxygen—gadolinium bonds.

Since there were no GdScO$_3$ reference available it was difficult to determine precisely the oxygen content, hence we will focus on the Sc/Gd ratio and assume that the layer is fully oxidized (the same oxidation process produced a fully oxidized Gd$_2$O$_3$ with an oxygen to gadolinium ratio of 1.51 ± 0.06). The peaks used to obtain the composition of the film were the Gd 4d and the Sc 2p, which resulted in a Sc/Gd ratio of 1.21 ± 0.03. Therefore, the chemical formula of the intermediate composition layer was Gd$_{0.9}$Sc$_{1.1}$O$_3$, indicating that the layer was a gadolinium scandate slightly Sc-rich, but close to stoichiometry.

Figure 3 represents the FTIR absorbance spectra from 1200 to 900 cm of the Gd$_{0.9}$Sc$_{1.1}$O$_3$ pure binary oxides. In the other parts of the spectrum there were no relevant absorptions. A slight peak centered at around 1040 cm$^{-1}$ related to substoichiometric SiO$_x$ [40, 41] is clearly observed for the pure Gd$_2$O$_3$ layer. The area of this peak can be related to the thickness of the SiO$_x$ at the interface. For the pure Sc$_2$O$_3$ and the Gd$_{0.9}$Sc$_{1.1}$O$_3$ films, this absorption is less intense. In both cases, the band is comparable in intensity and close to the detection limit. These spectra indicate that the regrowth of the interfacial oxide is not very significant for the as grown layers.

To study the evolution of the interface with the temperature treatments, in figure 4 it is presented the FTIR spectra for the Gd$_{0.9}$Sc$_{1.1}$O$_3$ film before and after the FGAs performed at 400 °C and 600 °C. In this figure it can be observed a mild increase of the area of the SiO$_x$ band as the annealing temperature raises. This behavior points out to the formation of Si–O bonds during the FGA, which have the consequence of a slight SiO$_x$ interlayer regrowth.

To study the atomic arrangement of the Gd$_{0.9}$Sc$_{1.1}$O$_3$ film, GIXRD diffraction measurements were performed after the FGA at 600 °C, and the results are represented in figure 5. No diffraction peaks were observed, pointing out to an amorphous layer, even after the FGA at 600 °C. Some works [6, 11, 42] reported for as deposited samples a peak located at ∼30° due to a Gd–Sc silicate formation. In those works, this peak disappeared after an annealing at 1000 °C. It is interesting to highlight that the sample shown in figure 5 does not present this peak, suggesting that whether there is no appreciable silicate layer or it is amorphous. As it was commented before, GdScO$_3$ is known to be a material with a good thermodynamic stability with Si up to 1000 °C. A typical microelectronic fabrication route will require back-end annealings under 600 °C. Thus, the result that the Gd$_{0.9}$Sc$_{1.1}$O$_3$ film remains in the amorphous phase at 600 °C is a desired property in order to avoid grain boundaries, that would increase the leakage current [43].

To confirm the amorphous character of the film the HRTEM image shown in figure 6 was obtained for the Gd$_{0.9}$Sc$_{1.1}$O$_3$ sample after the FGA at 600 °C. First of all, a stacked structure is observed, with two layers on top of the Si lattice. Both layers are amorphous (as it was shown with the GIXRD results of figure 5). The lighter layer is in contact with the Si and presents a thickness of 0.9 ± 0.1 nm. This layer corresponds to $SiO_x$, supporting the FTIR results from figure 4 which indicated a slight interlayer regrowth after the FGAs. Here, this regrowth can be quantified, and it is only 0.9 ± 0.1 nm even after the annealing at 600 °C, confirming the thermodynamical stability of gadolinium scandate on silicon. On top of $SiO_x$ a darker film is found, 5.0 ± 0.1 nm thick. This top layer is the amorphous gadolinium scandate. As in previous works [17], there are not traces of the presence of a nanolaminate, confirming the intermixing promoted by our fabrication process. The thickness of the film is fairly uniform for all analyzed regions, and there are no indications of poly-cristallinity.

Hence, from the physical characterization results it can be concluded that with HPS gadolinium scandate films close to stoichiometry can be grown. These films are amorphous and present a low regrowth of interfacial $SiO_x$ layer, even after a FGA at 600 °C.

From this point on we will focus on the electrical characteristics of MIS devices fabricated with $Gd_{0.9}Sc_{1.1}O_3$ as high-k dielectric. The area normalized capacitance C and conductance measured G at 10 kHz as a function of the gate voltage ($V_{gate}$) characteristics are depicted in figures 7(a) and (b) respectively, before and after representative FGAs. All the C–$V_{gate}$ curves are free of humps and do not present stretch out, indicating a good quality of the interface [31]. The as deposited sample (in gray solid line) presents the lowest accumulation capacitance. This value increases while the temperature of the FGA raises up to 600 °C. As the top electrode is Pt, which is a noble metal, this capacitance increase has to be associated to the gadolinium scandate formation. It was mandatory to produce the intermixing of Sc and Gd layers to form a homogeneous high-k gadolinium scandate. According to the C–$V_{gate}$ results of figure 7(a), this can be promoted by a low temperature treatment. The same effect was observed by our group in [17, 22] were we used binary $Gd_2O_3$ and $Sc_2O_3$ to form the ternary material after the FGA. Focusing on the G–$V_{gate}$ curves of figure 7(b), before the FGA, the conductance is higher than after the FGA at moderate temperatures. This fact indicates that at 300 °C, together with the formation of the gadolinium scandate, leakage paths are passivated. A slight increase in the conductance is observed as the FGA temperature is raised, most likely related to the accumulation capacitance increase commented in the former paragraph. For these temperatures, the conductance shows a peak in depletion, due to the interfacial traps response. In accumulation the conductance is moderate, lower than $10^{-1}$ s cm$^{-2}$ at a gate voltage of 2 V.

The effect of further increasing the FGA temperature up to 700 °C (presented in figure 7 with gray dashed lines) produces a severe capacitance roll-off due to a high conductance (at 0.5 V, the normalized conductance exceeds 1 S cm$^{-2}$). This increase in conductance is due to a severe leakage increase, that we will show in detail the following paragraphs. The capacitance roll-off is not a real decrease of gate capacitance but it is due to the limitations of the electrical probe station, which cannot measure accurately the capacitance in the presence of high conductance [44]. In any case, the high conductance at 700 °C suggests that the maximum FGA temperature that should be carried out for these films is 600 °C. The origin of this severe increase of conductance can be that the gadolinium scandate has reached the onset of crystallization, appearing grain boundaries that would act as leakage paths.

The EOT values for all FGA temperatures are represented in figure 8(a). We observe that the as deposited sample has an EOT of ~2.1 nm. This value decreases as the temperature of the FGA is raised. The EOT reaches a minimum value of ~1.5 nm after the FGA at 600 °C. At 700 °C the curve could not be fitted reliably since the accumulation region presented the capacitance roll-off. This EOT reduction after the FGA was not observed in the binary oxides, as it was found out in earlier works [17, 22, 23]. In fact, an increase in the EOT of around 0.2–0.5 nm for the binary oxides after the FGA was obtained, and the conclusion was that there was some interfacial regrowth. Here FTIR results showed that annealing temperature promoted some SiO$_x$ regrowth, thus the only explanation for EOT decrease is the permittivity boost of Gd$_{0.9}$Sc$_{1.1}$O$_3$. In other words, annealing promotes the nanolaminate intermixing and thus the gadolinium scandate formation together with its permittivity, higher than the binary constituents. A similar trend was found by Feijoo et al in [17].

Additionally, using the conductance method [31], it was found that the $D_{it}$ presents also a decreasing tendency with the temperature, as it is shown in figure 8(b). The permittivity boost does not compromise the interface quality, which is an excellent result. As expected, the $D_{it}$ value is reduced more than one order of magnitude after the first FGA (from $8 \times 10^{12}$ eV$^{-1}$ cm$^{-2}$ for the as deposited sample to $6 \times 10^{11}$ eV$^{-1}$ cm$^{-2}$ after the FGA at 300 °C). Then it further decreases slightly with temperature down to $3 \times 10^{11}$ eV$^{-1}$ cm$^{-2}$ for 600 °C. This $D_{it}$ value is similar to those reported in other works for GdScO$_3$ grown by electron beam evaporation [11] and ALD [45]. Besides, it is noteworthy that even at 600°C interface states do not depassivate (in other words, hydrogen remains bonded to defects). At 700 °C the trap density cannot be reliably assessed, but since the conductance peak is similar as the 600 °C case, no depassivation seems to occur neither.

It is important to highlight that the SiO$_x$ thickness found by HRTEM (figure 6) is very close to the EOT (that is 1.5 nm for the sample after the FGA at 600 °C). Thus, the ±0.1 nm uncertainty of the SiO$_x$ thickness will introduce a large variation when calculating the permittivity of the dielectric film. The effective k value of the dielectric stack can be obtained with the following equation:

$$k_{eff} = \frac{3.9t}{EOT}, \quad (1)$$

where t is the total dielectric thickness and 3.9 is the relative permittivity of SiO$_2$. However, if we want to obtain the permittivity value of the high-k material, the next formula has to be used:

$$k = \frac{3.9 t_{HK}}{EOT - t_{IL}} \quad (2)$$

being $t_{Hk}$, the thickness of the high-k dielectric and $t_{IL}$, the interlayer thickness. Therefore, using the thicknesses obtained in figure 6 and using equation (1), the $k_{eff}$ of the dielectric stack is higher than 15, in the same range as the HfSiO$_4$ reported in other works [46, 47] and used at present in the industry. Nevertheless, by means of equation (2), the permittivity of the Gd$_{0.9}$Sc$_{1.1}$O$_3$ film grown by HPS with the two-step method is 32, a value that is in the upper range of the reported values for this material, which is between 20 and 30 [4, 5, 6]. In fact, typical permittivities of amorphous GdScO are around 22–23 [11, 22, 48–50]. If we take into account the uncertainty of the interfacial $SiO_x$ thickness, the $Gd_{0.9}Sc_{1.1}O_3$ permittivity would be 28 ($t_{IL}$ = 0.8 nm) or even 39 ($t_{IL}$ = 1.0 nm), in all cases in the upper range of $GdScO_3$ permittivity. This excellent result confirms that the fabrication of MIS devices by means of HPS from metallic targets followed by a plasma oxidation is a very promising alternative for achieving a good performance gadolinium scandate.

To study the presence of slow traps, the hysteresis curves of these devices were measured before and after each FGA, starting the sweep from inversion to accumulation and back again. Figure 9 represents the hysteresis curves for three different FGAs: 300 °C, 500 °C and 600 °C. After the FGA at 300 °C (in the left hand side of this figure) the flatband voltage shift, $\Delta V_{FB}$, is small and clockwise. This is reduced with annealing temperature, as it can be seen on the sample annealed at 500 °C, which presents a negligible value. Finally, the FGA performed at 600 °C produces a counterclockwise (negative) hysteresis. Figure 10 shows the $\Delta V_{FB}$ value for all the annealing temperatures. There it can be appreciated that as deposited devices have an appreciable $\Delta V_{FB} \sim 150$ mV, that disappears when annealing at 350 °C. On the other hand, the $V_{FB}$ shift is noticeable again when annealing at 600 °C and above, but in this case with a negative value, around −160 mV. To explain this behavior, it is important to remark that the changes in the $V_{FB}$ are mainly determined by charge variations in the proximity of the oxide/semiconductor interface during the sweep. This oxide trapped charge, $Q_{ot}$, can be obtained using the following expression [51]:

$$Q_{ot} = -\Delta V_{FB} C_{ox} \quad (3)$$

where $C_{ox}$ is the oxide capacitance. The physical origin of this charge variation is either electrons from the semiconductor that are trapped by defects of the dielectric (clockwise or positive $\Delta V_{FB}$), or mobile positive ions (like sodium, potassium or hydrogen), that are pushed by the electric field from the gate towards the semiconductor (counterclockwise or negative $\Delta V_{FB}$). Thus the sign and value of $\Delta V_{FB}$ gives insight into the dielectric quality.

A possible explanation of the results shown in figure 10 can be that as deposited $Gd_{0.9}Sc_{1.1}O_3$ film presents some dangling bonds that act as electron traps, and thus produce a clockwise $\Delta V_{FB}$. These traps are passivated at 350 °C by the hydrogen from the forming gas atmosphere and as a consequence, hysteresis disappears. At 600 °C, the counterclockwise flatband voltage shift appears, indicating a displacement of positive ions with polarity. Since no mobile ions are present at lower temperatures, K or Na contamination at 600 °C is not likely. Therefore, the most plausible candidate is excess hydrogen from the FGA. This means that at 600 °C and above an excess of hydrogen is accumulated within the dielectric and gives rise to the negative $\Delta V_{FB}$.

The C–$V_{gate}$ and G–$V_{gate}$ frequency dispersion characteristics measured from 1 kHz to 1 MHz are represented in figure 11 for the $Gd_{0.9}Sc_{1.1}O_3$ sample after the FGA at 600 °C. No frequency dispersion of the flatband voltage is found, and the hump in depletion due to interface traps is minimal. The only appreciable effect in the C–$V_{gate}$ curves is a reduction in the accumulation capacitance measured at 1 MHz, which is originated from the coupled effect of high conductance and substrate series resistance. This last parameter mainly affects the high frequency C–$V_{gate}$ curve [52]. The value of the series resistance for the 1 MHz curve is ~170 Ω. From these curves and using the conductance method [31], the $D_{it}$ is around $3 \times 10^{11}$ eV$^{-1}$ cm$^{-2}$ for all frequencies, pointing out that the interfacial traps are fast and can respond even at high frequencies. Analogous results were found for the other annealing temperatures.

The leakage of the $Gd_{0.9}Sc_{1.1}O_3$ films measured before and after different FGAs is presented in figure 12. The as grown sample has one order of magnitude higher leakage current than the sample with a FGA at 300 °C. As it was commented in the C–$V_{gate}$ curves of figure 7, the formation of the gadolinium scandate required a temperature treatment. Here this conclusion is also confirmed: the FGA reduces the current density due to leakage paths passivation. For the FGA at 300 °C the leakage at a gate voltage of 0.3 V is $5 \times 10^{-7}$ A cm$^{-2}$, which is close to the roadmap target ($1 \times 10^{-7}$ A cm$^{-2}$). In the subsequent annealings the leakage current increases as the temperature of the FGA is raised, together with the reduction in the EOT observed before. This points out to a densification of the $Gd_{0.9}Sc_{1.1}O_3$ layer that increases the tunneling current. The current density after the FGA at 600 °C is moderate $2 \times 10^{-3}$ A cm$^{-2}$ ($V_{FB}$ + 1.5) V for an EOT of 1.5 nm. This value is more than four orders of magnitude lower compared with the current of a capacitor with an equivalent 1.5 nm $SiO_2$ layer [53].

Thus, the electrical results show that $Gd_{0.9}Sc_{1.1}O_3$ grown by the two-step method produces MIS devices with good electrical behavior, obtaining an EOT of ~1.5 nm with low interface traps density, hysteresis, leakage current, and a very high permittivity of ~32. The optimal FGA temperature is found to be in the 500 °C–600 °C range: 500 °C is better from a leakage and hysteresis standpoint, but 600 °C has lower EOT and interface trap density.

However the 0.9 nm thick interfacial layer could limit an aggressive scaling of the EOT. A process that is being researched to reduce this interfacial layer is the remote scavenging, as explained in the introduction. Here, as a proof of concept for HPS deposited $GdScO_3$ we tested the scavenging potential of Ti electrodes. Therefore, a Ti/Pt stack (with thicknesses of 5 nm and 25 nm, respectively) was evaporated after a 5 min FGA of the $Gd_{0.9}Sc_{1.1}O_3$ at 600 °C to ensure the formation of the gadolinium scandate before the scavenging process. After Ti evaporation, a FGA at 300 °C for 20 min was performed to improve the backside metallic contacts and produce scavenging. The scavenging electrode and FGA conditions used where the best ones that were optimized in [27] for 4 nm of $Gd_2O_3$.

In figure 13 we represent the C–V$_{gate}$ for the capacitors with $Gd_{0.9}Sc_{1.1}O_3$ as dielectric. The Ti-gated devices (solid line) present a significant increase in the accumulation capacitance compared to the Pt samples after the FGA at 600 °C (shown as a reference by the gray line) from 1.9 to ~2.25 μF

cm$^{-2}$. This increase is due to the remote scavenging effect produced by the Ti layer. It is important to remember that in these devices, the interface layer is only 0.9 nm thick and thus, a small reduction in the thickness of this value would clearly increase the capacitance. Additionally, an accumulation capacitance roll-off is observed for this Ti sample at $V_{gate} > 1.0$ V. This is due to the high conductance, that is higher than $2 \times 10^{-1}$ S cm$^{-2}$, two orders of magnitude higher than in the Pt case. The EOT value is around 1.2 nm for this Ti-gated sample before and after annealing. Besides, a difference of around 0.9 V is observed in the $V_{FB}$ for the Pt and Ti samples, which is in accordance with the workfunction difference of these metals [31].

Furthermore, scavenging the interface has the detrimental effect of severely increasing the interface trap density: for the Ti devices the $D_{it}$ is high ($\sim 10^{13}$ eV$^{-1}$ cm$^{-2}$) due to the formation of additional dangling bonds, as has been reported in other works on scavenging [25, 54, 55]. This value is almost two orders of magnitude higher than the Pt case, which was $3 \times 10^{11}$ eV$^{-1}$ cm$^{-2}$.

Figure 12 also depicts the leakage current density versus gate voltage characteristics for the devices with Gd$_{0.9}$Sc$_{1.1}$O$_3$ after Ti scavenging. The Ti sample increases the leakage current with respect to the Pt device ($10^{-2}$ A cm$^{-2}$ and $\sim 2 \times 10^{-3}$ A cm$^{-2}$, respectively at $V_{FB} + 1.5$ V). This is another known effect of scavenging: since the SiO$_2$-like interfacial layer thickness is reduced (or even disappears), the tunneling probability is enhanced, and also the oxygen displacement leaves dangling bonds that can act as hopping centers [56]. In any case, the main conclusion of this section is that the compatibility of the scavenging effect with gadolinium scandate formed by our optimized two-step HPS process has been demonstrated. Further work will be devoted to characterize this scavenging effect aiming at its optimization.

4. **Summary and conclusions**

In this article the fabrication of gadolinium scandate films was achieved by means of HPS after a deposition of a nanolaminate of metallic Gd and Sc layers and a subsequently plasma oxidation. A gadolinium scandate with a composition close to stoichiometry was obtained. This Gd$_{0.9}$Sc$_{1.1}$O$_3$ film was amorphous and presented good electrical characteristics: for an EOT of 1.5 nm, the leakage current, the hysteresis and the interfacial trap density were low and similar to values reported in other works. The k value of this dielectric film is around 32, which is among the highest reported values. The low EOT is mostly determined by the interface layer thickness (0.9 nm after the FGA at 600 °C), so decreasing this interface is crucial for further EOT scaling. Therefore, the proven compatibility of this material with the scavenging effect paves the path for further improving the EOT.

In the future we will focus on the leakage reduction by extending the FGA time and/or varying the annealing atmosphere at moderate temperatures, and by systematically studying the plasma oxidation conditions to ensure complete GdScO oxidation. Also, we will study the compatibility of HPS deposited GdScO with high mobility substrates, such as InP, Ge or InGaAs.

**Acknowledgments**

The authors acknowledge the 'CAI de Técnicas Físicas', the 'CAI de Espectroscopía', 'CAI de Difracción de Rayos X' and 'Centro Nacional de Microscopía Electrónica' of the Universidad Complutense de Madrid for sample fabrication, FTIR, GIXRD and HRTEM measurements, respectively. This work was funded by the project TEC2010-18051 from the Spanish 'Ministerio de Economía y Competitividad', the 'Formación de Personal Investigador' program under grant BES-2011-043798, and the 'Juan de la Cierva—Formación' program under grant FJCI-2014-19643.

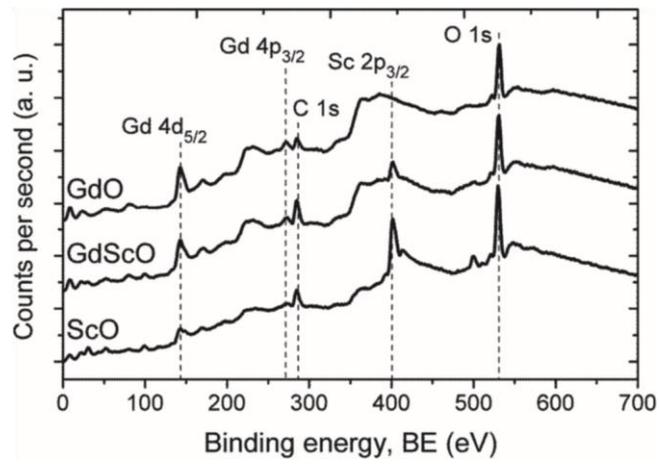

Figure 1. XPS wide scan or survey spectra of $Gd_2O_3$, $Gd_{2-x}Sc_xO_3$ and $Sc_2O_3$ samples after a FGA at 600 °C. $Gd\,4d_{5/2}$, $Sc\,2p_{3/2}$ and $O\,1s$ peaks are marked in the figure.

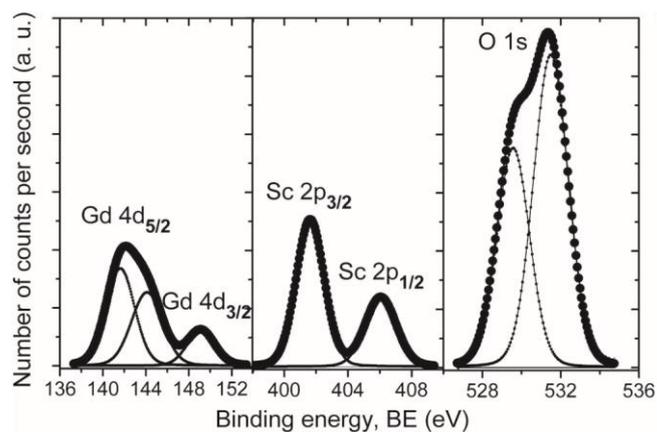

Figure 2. High resolution XPS spectrum for $Gd_{2-x}Sc_xO_3$ sample after a FGA at 600 °C: Gd 4d (left) and Sc 2p (center) doublets and O 1s (right) together with their fits.

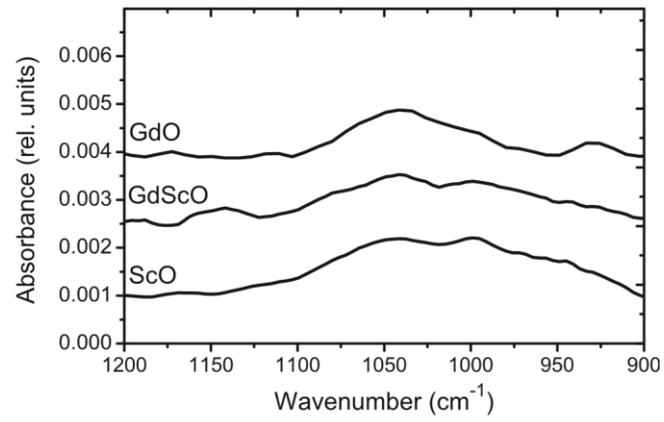

Figure 3. FTIR absorbance spectra for the as deposited $Gd_2O_3$, $Gd_{0.9}Sc_{1.1}O_3$ and $Sc_2O_3$ samples. Curves are displaced vertically for the sake of clarity.

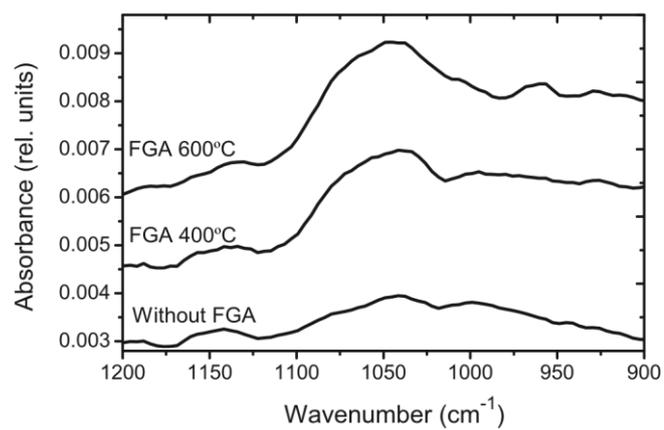

Figure 4. FTIR spectra for the $Gd_{0.9}Sc_{1.1}O_3$ film before and after different FGAs. Curves are displaced vertically for the sake of clarity.

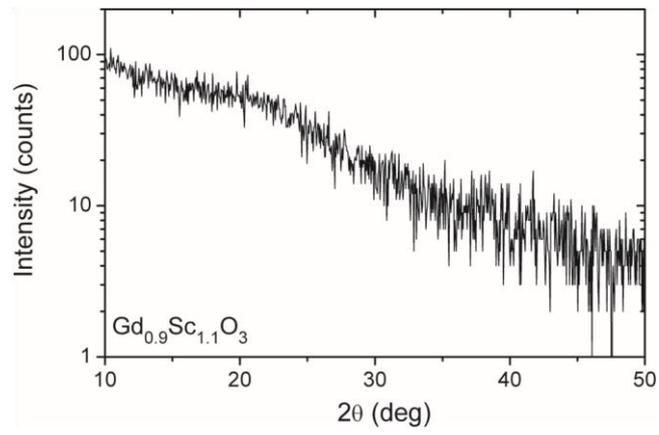

Figure 5. GIXRD spectrum of $Gd_{0.9}Sc_{1.1}O_3$ film after a FGA at 600 °C was performed. No diffraction peaks are observed.

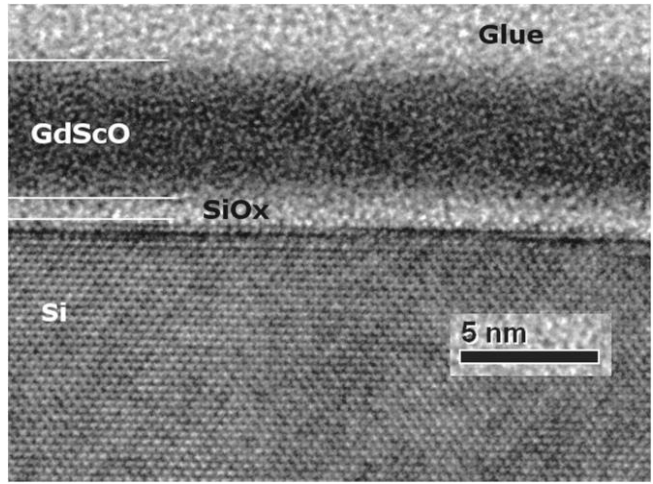

Figure 6. HRTEM cross-sectional image of $Gd_{0.9}Sc_{1.1}O_3$ sample obtained after the sequential FGA up to 600 °C.

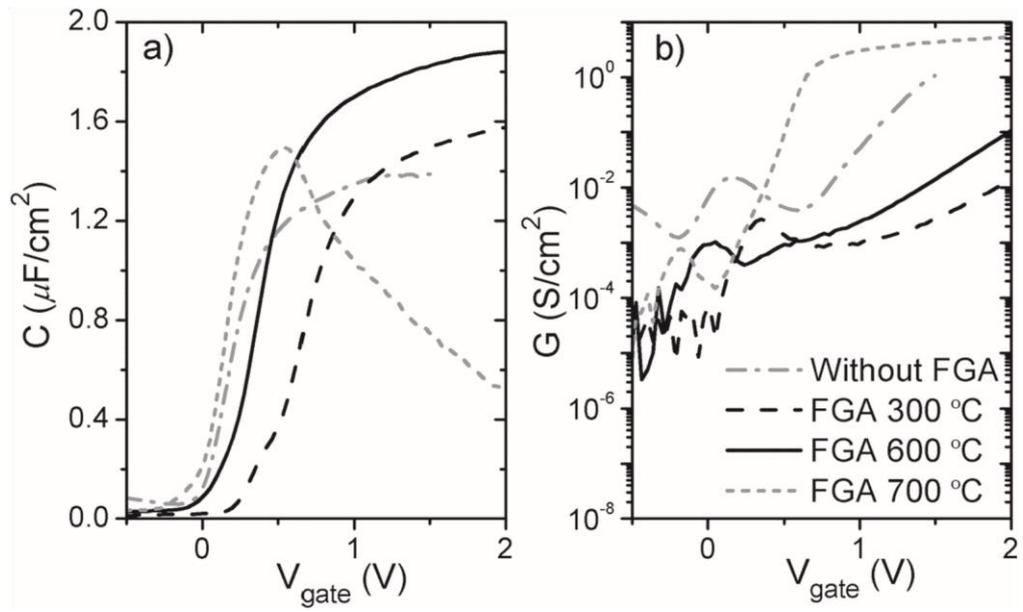

Figure 7. Area normalized capacitance and conductance as a function of $V_{gate}$ characteristics for the Pt-gated $Gd_{0.9}Sc_{1.1}O_3$ sample before and after representative FGAs.

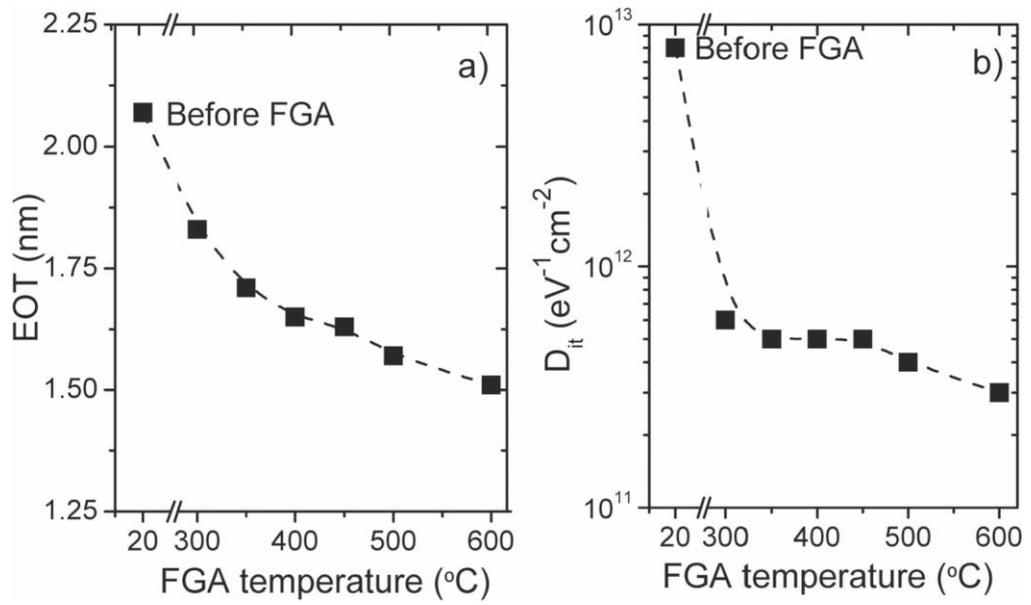

Figure 8. (a) EOT and (b) $D_{it}$ as a function of the annealing temperature for the sample fabricated with $Gd_{0.9}Sc_{1.1}O_3$ as dielectric.

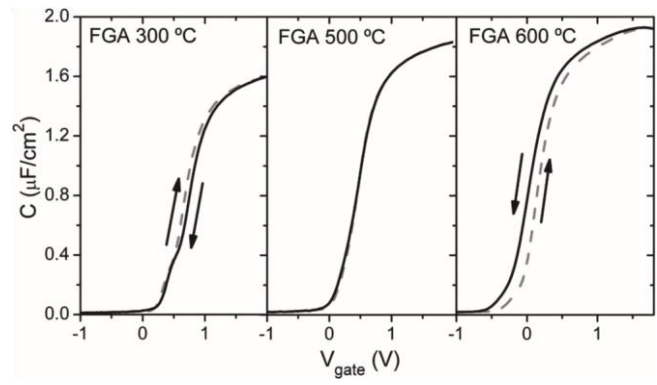

Figure 9. Hysteresis C–$V_{gate}$ characteristics measured from inversion to accumulation and back again for the $Gd_{0.9}Sc_{1.1}O_3$ sample after several FGAs at different temperatures.

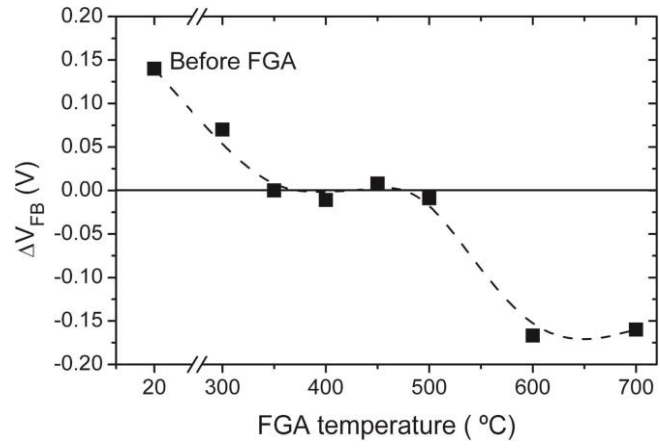

Figure 10. Flatband voltage shift as a function of the annealing temperature for the Pt-gated Gd$_{0.9}$Sc$_{1.1}$O$_3$ sample.

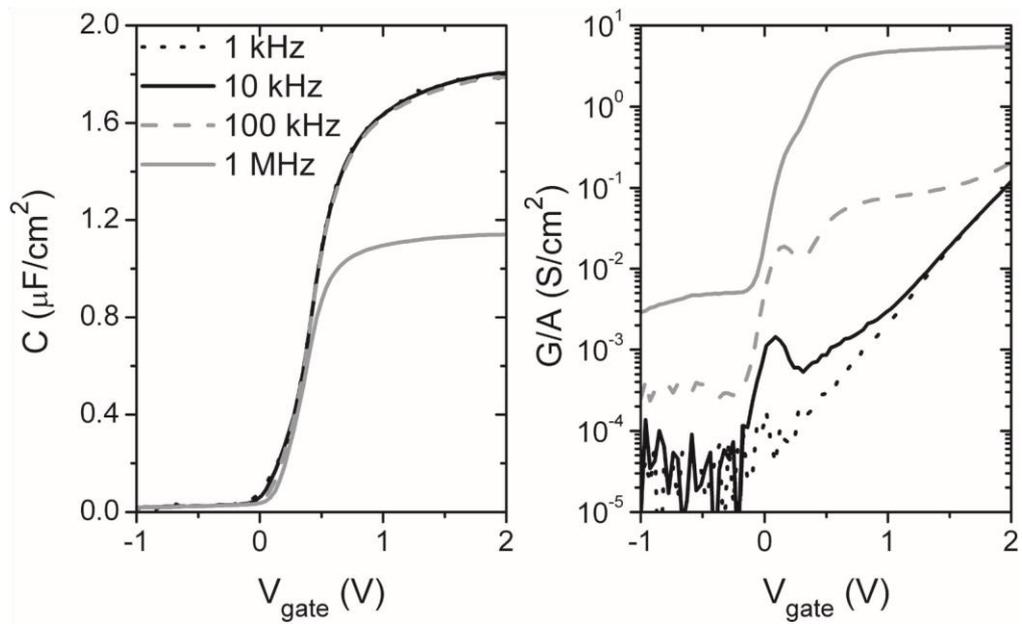

Figure 11. C–V$_{gate}$ and G–V$_{gate}$ curves measured at different frequencies for the Gd$_{0.9}$Sc$_{1.1}$O$_3$ sample after FGA at 600 °C.

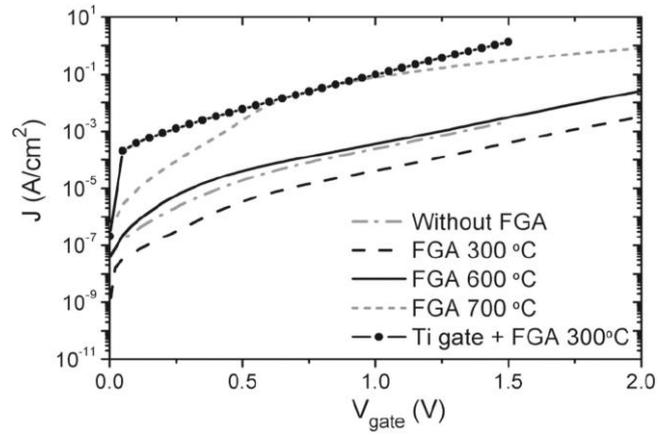

Figure 12. J–V$_{gate}$ characteristics for the Pt/Gd$_{0.9}$Sc$_{1.1}$O$_3$/Si devices before and after different FGAs (lines) and Pt/Ti/Gd$_{0.9}$Sc$_{1.1}$O$_3$/Si after FGA at 300°C (circles + line).

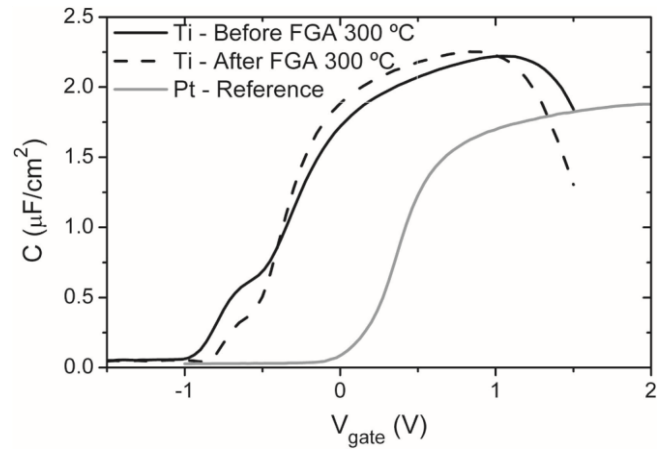

Figure 13. C–V$_{gate}$ characteristics of capacitors with Gd$_{0.9}$Sc$_{1.1}$O$_3$ formed after a FGA at 600 °C using Ti as top electrode (in black) before (solid lines) and after a second FGA at 300 °C (dashed lines). The same dielectric with Pt is represented as a reference in gray.